\documentstyle[graphicx,multicol,prb,aps]{revtex}
\begin{document}
%\draft
\title{Antiferromagnetic long-range order in Cu$_{1-x}$Zn$_x$GeO$_3$ with
extremely low Zn concentration
}
\author{K. Manabe,${}^{1,}$\cite{Manabe-address} 
H. Ishimoto,${}^1$
N. Koide,${}^{2,}$\cite{Koide-address}
Y. Sasago,${}^{2,}$\cite{Sasago-address} 
and K. Uchinokura${}^{2,}$\cite{Uchinokura-email}
}%
\address{${}^1$Institute for Solid State Physics, The University of Tokyo,
7-22-1 Roppongi,  Minato-ku, Tokyo 106-8666, Japan\\
${}^2$Department of Applied Physics, The University of Tokyo, 7-3-1
Bunkyo-ku, Tokyo 113-8656, Japan}
\date{March 24, 1998}
\begin{multicols}{2}[
\maketitle
%abstract%
\begin{abstract}
 We have measured the magnetic susceptibilities of single crystals
of Cu$_{1-x}$Zn$_x$GeO$_3$ with 
extremely low Zn concentration ($x$) lower than $x=5\times10^{-3}$
at very low temperatures to investigate the
spin-Peierls and antiferromagnetic
transitions.
The results show
that the undoped CuGeO${}_3$ has no antiferromagnetic phase down to
12 mK and there exists an antiferromagnetic long-range order with the
easy axis along the $c$ axis for
$x$ down to $1.12(2)\times10^{-3}$. 
The minimum observed
N\'eel temperature was 0.0285 K for $x= 1.12(2)\times 10^{-3}$ sample.
From the concentration dependence of the N\'eel temperature it
is concluded that there is no critical concentration
for the occurrence of the antiferromagnetic long-range order.
This indicates that the dimerization sustains the coherence of the 
antiferromagnetic phase of the spin polarization in impurity-doped systems
and is consistent with the theory
of the impurity-doped spin-Peierls system.
The temperature dependence of the susceptibilities
at $T>T_N$ of all samples indicates that the magnetic correlations between
localized spins are enhanced by
a relatively large interchain interaction
of CuGeO${}_3$.
\end{abstract}
]

\narrowtext
% body of paper here
Low-dimensional antiferromagnetic (AF) systems are very
attractive because they exhibit  various
interesting phenomena such as spin-Peierls (SP)
transition,\cite{Bray75} Haldane gap,\cite{Haldane83}
high-temperature superconductivity.\cite{Bednorz86}
Among them, the SP transition occurs in one-dimensional
AF spin systems with spin half ($S=1/2$),
%(generally speaking, with spin half odd integer at least theoretically),
whose ground state has a spin gap as a result of a
dimerization of the chain.
This transition drew much attention by a recent discovery of
the first inorganic SP system CuGeO$_3$  by Hase, Terasaki and
Uchinokura.\cite{Hase93a}
The compound has an  orthorhombic
crystal structure \cite{Vollenkle67}
which consists of linear chains of Cu$^{2+}$ ions ($S=1/2$)
along the $c$ axis, well separated from one another
by Ge-O chains but still weakly coupled antiferromagnetically
along the $b$ axis and ferromagnetically along the $a$
axis.\cite{Nishi94}
This cuprate has a great advantage for the study of the
SP transition over the organic SP materials, because we can easily
study the effect of substitution on the SP
transition.\cite{Hase93b}
Up to now the substitutions  by Zn$^{2+}$
($S=0$)\cite{Hase95a}
and Ni$^{2+}$ (S = 1)\cite{Koide96} for Cu$^{2+}$
 and by Si$^{4+}$ for Ge$^{4+}$ (Refs.~\onlinecite{Renard}
and \onlinecite{Regnault}) were performed
in detail,
resulting into the suppression of the SP transition and
the occurrence of the antiferromagnetic long-range order (AF-LRO).
Especially, in Cu$_{1-x}$Zn$_x$GeO$_3$,
a neutron scattering experiment showed a clear evidence for the
coexistence
of AF-LRO and lattice dimerization.\cite{Sasago96a,Martin97}
The coexistence is very interesting
because these two types
of LRO's were generally believed to be
mutually exclusive.
However Fukuyama {\it et al.}
 proposed a theoretical
model of impurity-induced AF-LRO in a SP system (in particular
for CuGe${}_{1-x}$Si${}_x$O${}_3$).\cite{Fukuyama96a}

At present the AF-LRO is believed to be induced by the
substitution for Cu${}^{2+}$, which has $S=1/2$ spin on it, by other
ions:  Zn${}^{2+}$ ($S=0$) or Ni${}^{2+}$ ($S=1$) {\it etc.}
(or Ge${}^{4+}$ ($S=0$) by Si${}^{4+}$ ($S=0$)
in the case of CuGe${}_{1-x}$Si${}_x$O${}_3$).
In this case we may speculate that there exists a critical
concentration $x_c$ below which AF-LRO does not occur
even at $T=0$ K.
An alternative possibility is that AF-LRO persists to zero
impurity concentration (no critical concentration).
Anyway it is very interesting to see how extremely dilute
substitution causes or affects the AF-LRO.
Very recently Grenier {\it et al.}\cite{Grenier98}\ have performed
the measurements on
the susceptibility of the single crystals of CuGe${}_{1-x}$Si${}_x$O${}_3$
with small Si concentration ($x \geq 2\times10^{-3}$) down to 70 mK
and observed the occurrence of the AF-LRO.
But they failed to confirm whether critical concentration
exists or not because of the uncertainty of Si concentration
in low Si-doping level (according to them $x_{\mbox{\scriptsize Si}}$
corresponds to
$x_{\mbox{\scriptsize Zn}}/3$, see the discussion section).

In the present  work, we extended
the measurements on  the magnetic susceptibility of Cu$_{1-x}$Zn$_x$GeO$_3$
single crystals down to extremely
low Zn concentration of  $x=1.12\times10^{-3}$
to determine
the behavior of the N\'eel temperature in
that region.
For that purpose we have also extended the temperature region
down to 5 mK.

A series of Cu$_{1-x}$Zn$_x$GeO$_3$
single crystals were grown
using a  floating-zone method.
No traces of other impurity phases were detected in x-ray diffraction
patterns.
The Zn concentration $x$  was determined
with Inductively-Coupled-Plasma Atomic Emission Spectroscopy (ICP-AES)
 and Electron Probe Micro Analyzer (EPMA) for $x < 5\times 10^{-3}$ and
 $x > 5\times 10^{-3}$, respectively.
The samples for the ultra-low temperature measurements were grown,
doped with $x =1.12(2)$, 1.18(2), 2.13(4), 3.08(6), 3.81(8), 4.91(10),
 7(2), 13(2), 17(3), 18(3), 21(2), 22(2), 24(3), 26(2)$\times 10^{-3}$ Zn,
where the values in the parentheses are
experimental errors, or undoped.

%%%%%%%%%%%%%%%%%%%%%%%%%%%%%
\begin{figure}[t]
\begin{center}
\includegraphics*[width=7cm]{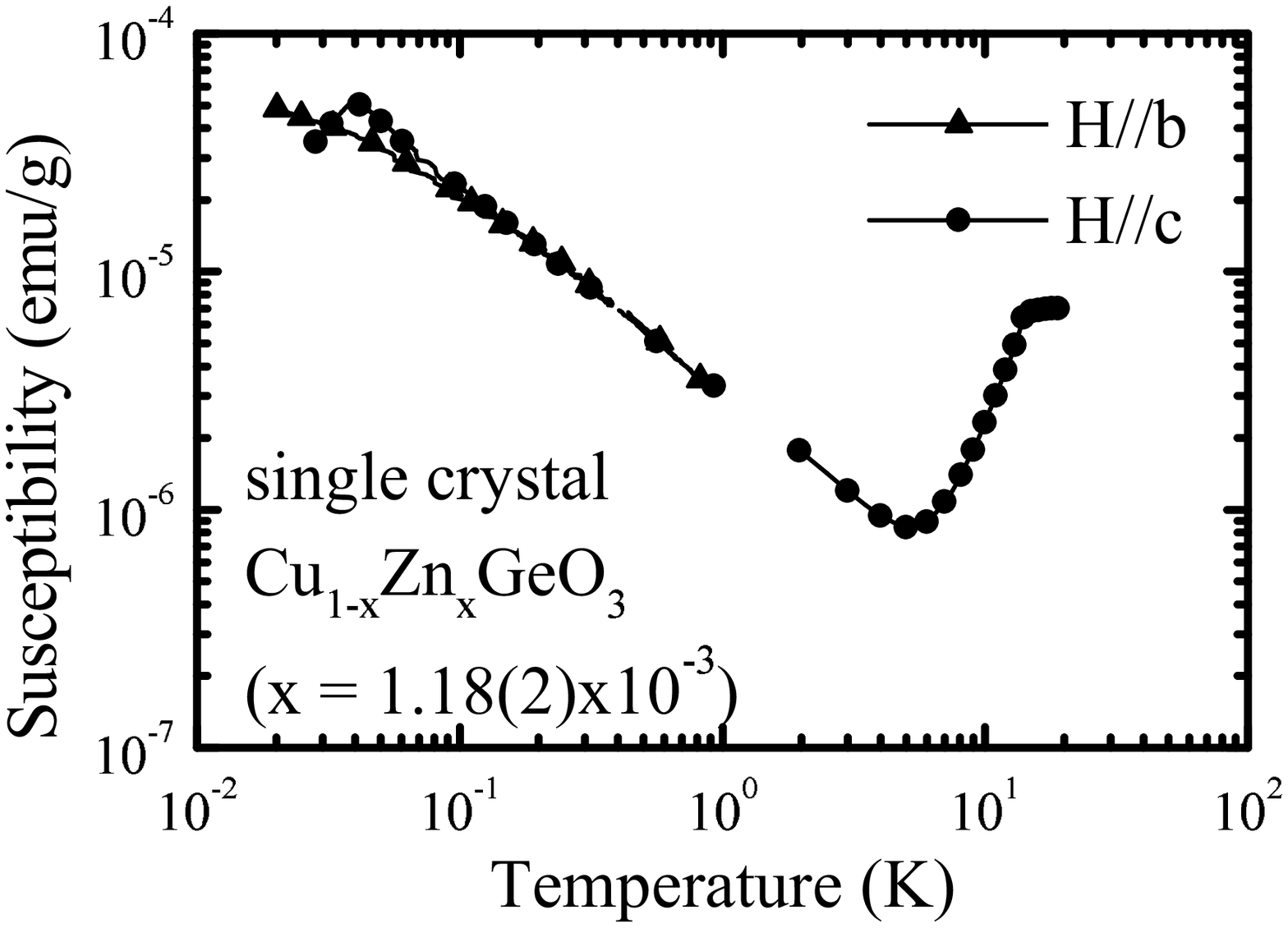}
\end{center}
\caption{%
The temperature dependence of $\chi(T)$ of
the single-crystal Cu$_{1-x}$Zn$_x$GeO$_3$
with $x = 1.18(2)\times10^{-3}$.
The  solid curves represent many experimental points for simplicity.
}
\label{fig1}
\end{figure}
%%%%%%%%%%%%%%%%%%%%%%%%%%%%%%%%%%

For the measurements below 2 K each sample was stuck on a quartz glass 
with a small amount of N-grease
to reduce a background signal. 
It was immersed in  nonmagnetic liquid ${}^4$He which was cooled by way 
of a sintered powder heat exchanger with a combination of  ${}^3$He-${}^4$He
dilution refrigerator and PrNi${}_5$  adiabatic demagnetization 
refrigerator. 
The temperature was determined with the germanium
and carbon resistor thermometers  above 20 mK and with a platinum NMR
thermometer below 20 mK, both  calibrated against the $^3$He melting curve.
Thermal contact between the sample and the coolant was found good enough
since  the susceptibility of the sample quickly  followed the
temperature change of the refrigerator and showed no hysteresis on cooling
and warming processes above 12 mK.
AC susceptibility was measured with a SQUID (Superconducting
QUantum Interference Device)
magnetometer where the amplitude and the frequency
of the applied AC field were $5\times10^{-3}$ mT and 16 Hz,
respectively.
To determine SP transition temperatures for all samples,
DC susceptibility measurements were also performed  at H = 0.1 T above
2 K in a separate cryostat.

\begin{figure}[t]
\begin{center}
\includegraphics*[width=8cm]{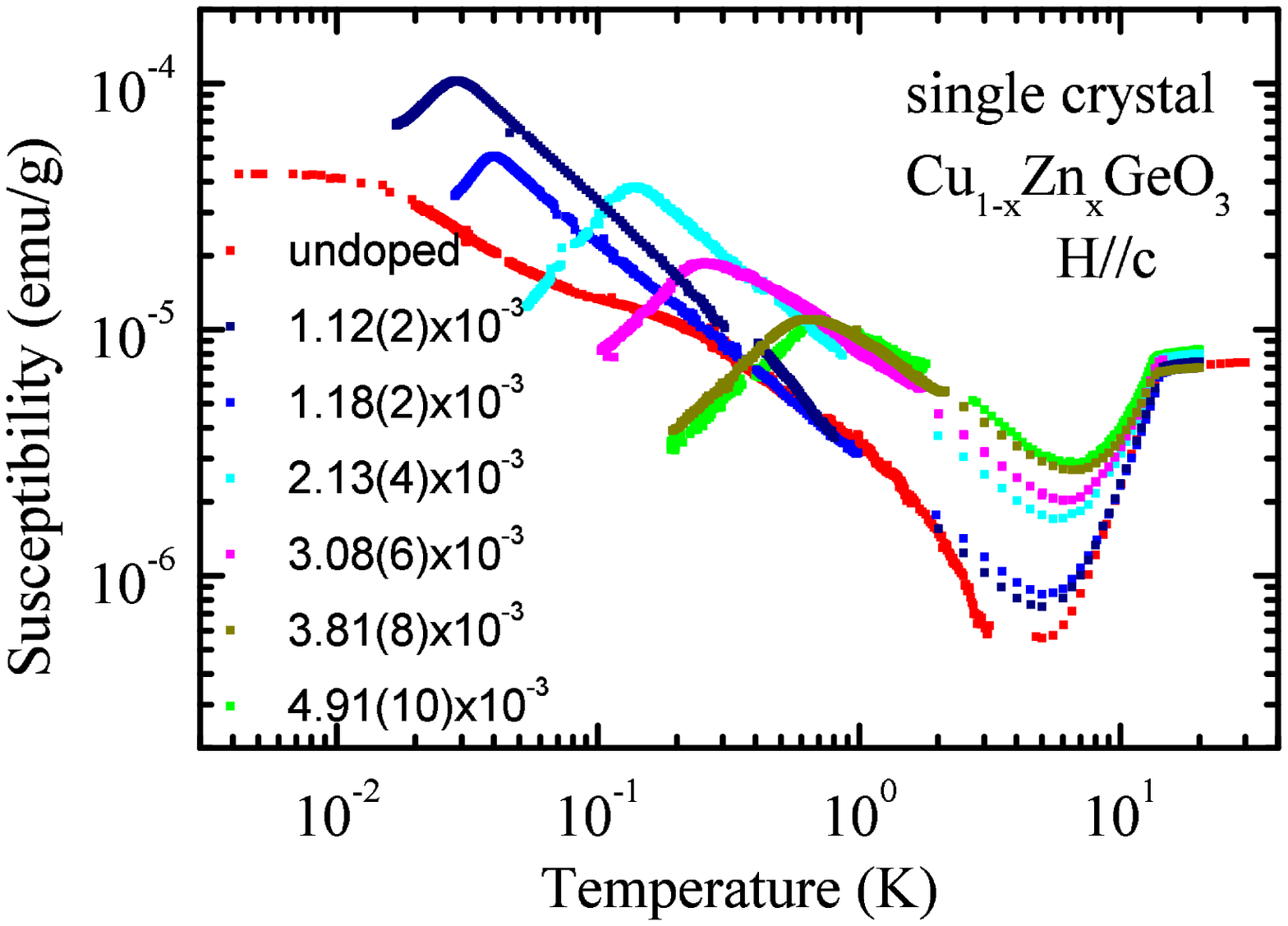}
\end{center}
\caption{(color)
The temperature dependence of $\chi_c(T)$ of the
single-crystal Cu$_{1-x}$Zn$_x$GeO$_3$ with
$x<5\times10^{-3}$ in AC magnetic field.
Numerals denote the Zn concentrations $x$.
}%
\label{fig2}
\end{figure}

The temperature dependence of the susceptibilities
along the $b$  ($\chi_b(T)$) and
$c$ ($\chi_c(T)$) axes  for Cu$_{1-x}$Zn$_x$GeO$_3$
with $x = 1.18(2)\times10^{-3}$ is shown in Fig.~\ref{fig1}.
Two phase transitions are clearly seen in $\chi_c(T)$.
One is characterized by a rapid drop at around 14.3 K,
corresponding to the suppressed SP transition by
the Zn substitution, which was first reported in
Ref.~\onlinecite{Hase93b}.
The other shows a cusp in $\chi_c(T)$ at around 0.04 K, below which
$\chi_c(T)$ decreases toward zero
whereas $\chi_b(T)$ increases slightly.
The anisotropic behavior clearly  indicates that
the transition at 0.04 K is the AF (N\'eel)
transition with the easy axis parallel to the $c$ axis,
as was observed in more heavily Zn-doped samples.\cite{Hase95a}
A similar behavior was observed in the sample with
$x=4.91(10)\times10^{-3}$, so only $\chi_c(T)$ was measured
for the other samples.
The results for $\chi_c(T)$ in the
low concentrtion region are summarized in Fig.~\ref{fig2}.
All the samples exhibit both the SP
and the AF transitions  around 14 K and below 1 K, respectively.

%%%%%%%%%%%%%%%%%%%%%%%%%%%%%
%%%%%%%%%%%%%%%%%%%%%%%%%%%%%%%%%%%

\begin{figure}
\begin{center}
\includegraphics*[width=7cm]{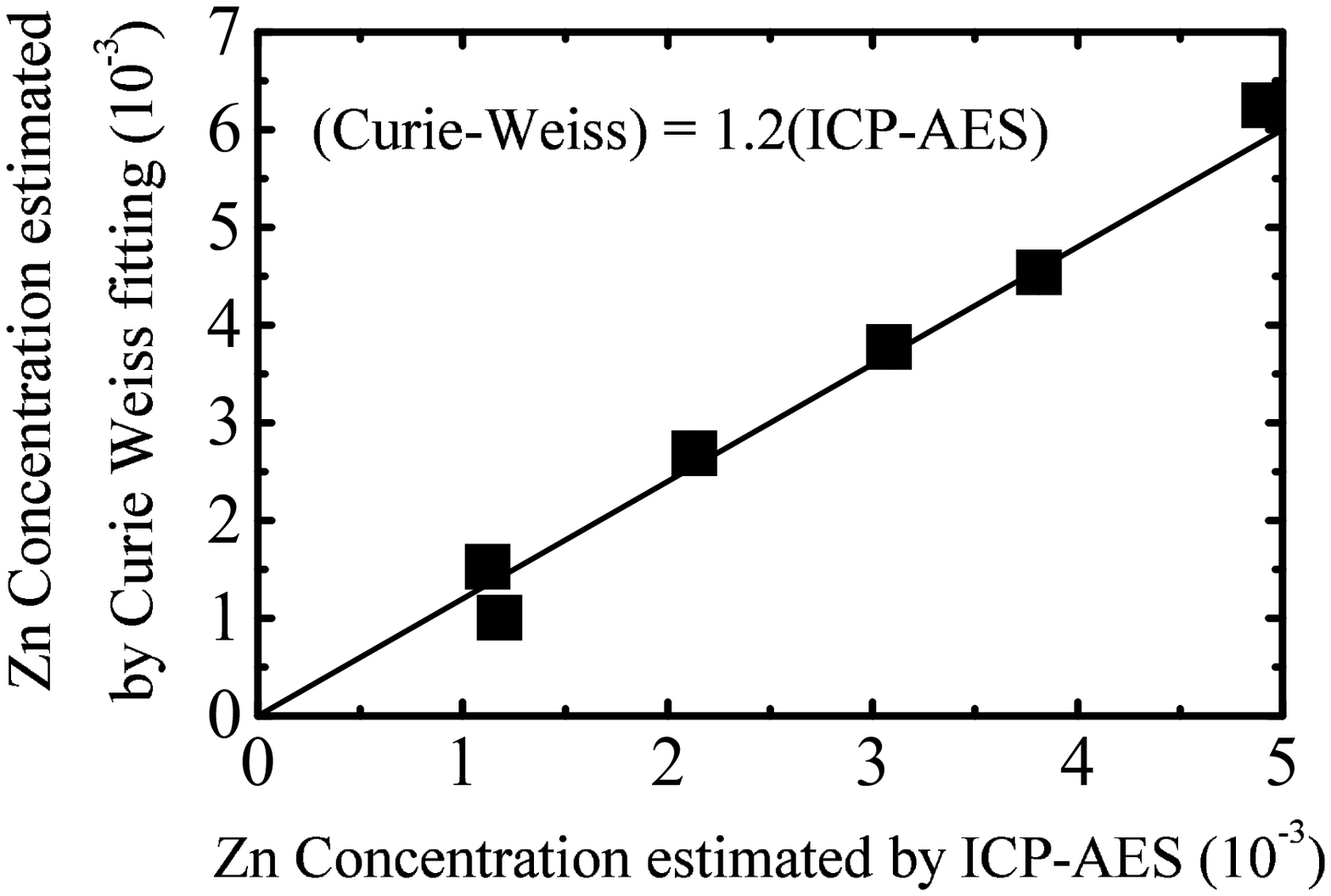}
\end{center}
\caption{%
Relation between Zn concentrations estimated by ICP-AES and those
estimated by Curie-Weiss fitting.
}%
\label{fig3}
\end{figure}

In the same figure, the result is also given  for the undoped
sample which was prepared in the same way as the doped ones.
Although the concentration of impurities is extremely low
in the undoped sample,
it may contain defects which play the same role as
Zn$^{2+}$ ions.
Both the impurities and defects cut the  dimerized chains of Cu$^{2+}$  
and produce nearly isolated 
$S=1/2$ states on the broken edges in the case of large distance
between defects.
Thus such defects (or Zn$^{2+}$)  cause the increasing
susceptibility  in the spin-Peierls state and we can estimate their
concentration by the Curie-Weiss fitting.
In Fig.~\ref{fig3} we show the relation between Zn concentration
estimated by ICP-AES and those estimated by Curie-Weiss fitting
in low Zn concentration region ($x<5\times 10^{-3}$).
In spite of extremely low Zn concentration,
the latter agrees with the former within an accuracy of 20\%.
This shows the validity of the determination of Zn concentration.
The fitting for the undoped sample gives us the effective value of
$x = 2.3(2)\times10^{-4}$.
Its susceptibility seems to saturate below 0.012 K.
There are two possible reasons for
the occurrence of this saturation.
One is that the undoped sample was not cooled below
0.012 K.
The other is that the susceptibility is indeed constant
below 0.012 K, which indicates that the AF transition may occur 
below 0.012 K.
We cannot determine which is the case.
But we can say that above 0.012 K AF transition does not occur
in the undoped sample.

\begin{figure}[t]
\begin{center}
\includegraphics*[width=8cm]{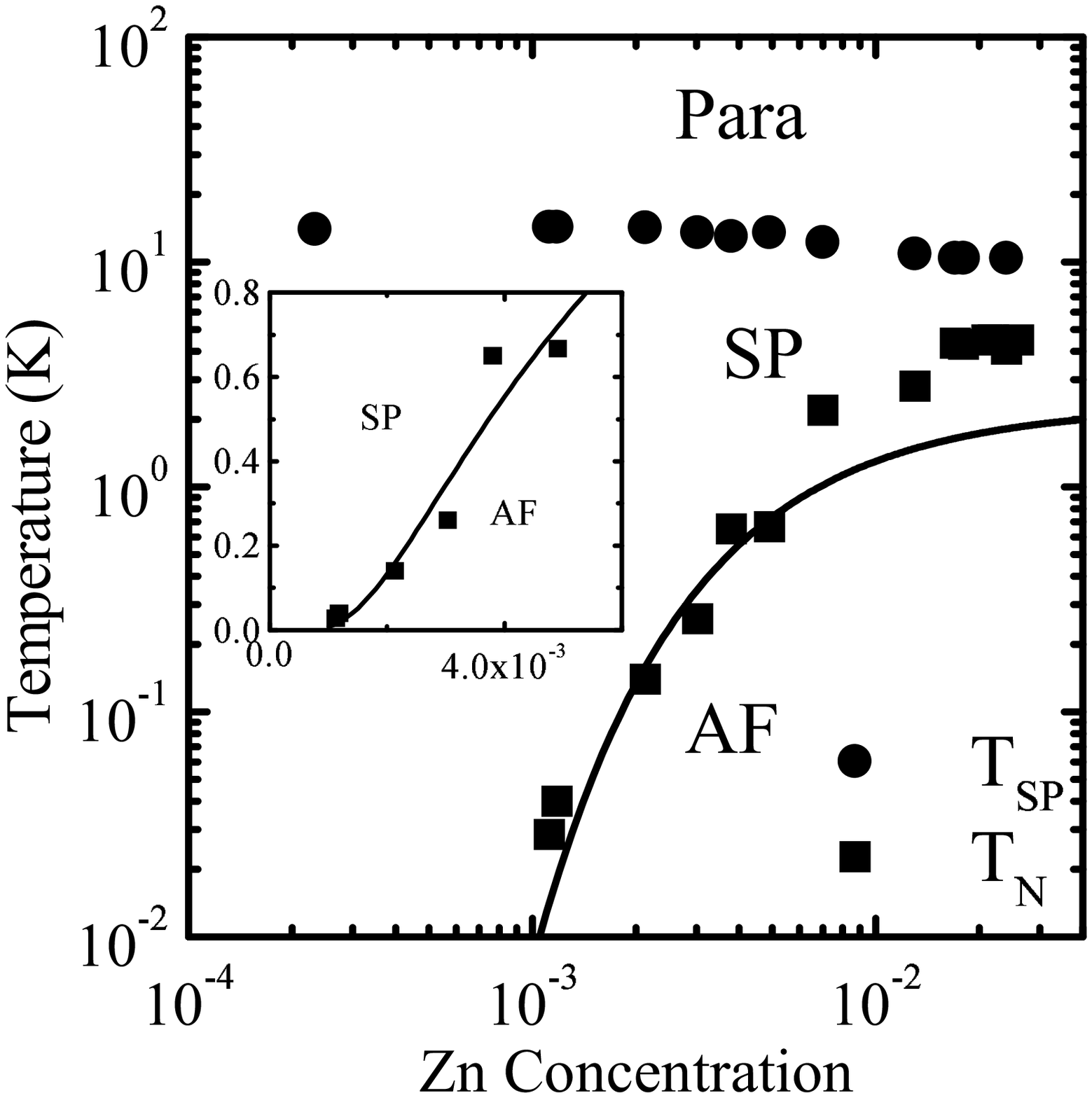}
\end{center}
\caption{%
$T_N$ and $T_{SP}$ as
a function of Zn concentration $x$.
   Para, SP and AF represent paramagnetic,
   spin-Peierls and antiferromagnetic states, respectively.
   Solid curve is the best fitted one by Eq.~(\ref{eq.Tn}).
The inset shows the same relation in the low concentration region
in linear scales.
}%
\label{fig4}
\end{figure}

The  N\'eel temperature $T_N$  and SP transition temperature
$T_{SP}$ are summarized in Fig.~\ref{fig4} for all samples.
The behavior of $T_N$ in very low concentration region and
the appearance of the AF transition at 0.0285 K for such low
Zn concentration as
$x = 1.12(2)\times10^{-3}$ imply that there is no critical concentration
for the occurrence of the AF-LRO.
This may come from the fact that the dimerization sustains
the phase coherence of the spin polarization although it suppresses
the magnitude of the spin polarization.

%%%%%%%%%%%%%%%%%%%%%%%%%%%%%
%%%%%%%%%%%%%%%%%%%%%%%%%%%%%%%%%%%%%%%%

The absence of critical concentration is also consistent with the
the theory of the impurity-doped SP system.\cite{Fukuyama96a}
According to this theory, no matter how low the concentration of
the impurity is,
it causes the AF-LRO because SP state has the LRO of the dimerization.

In the low Zn concentration ($1.12(2)\times10^{-3} \leq x \leq
4.91(10)\times10^{-3}$) $T_{N}$ can be fitted to a
formula:
%%%%%%%%%%%%%%%%%%%%%%%%%%%%%%%%%%%%%%%%%%%%%%%%%%%%%%%
\begin{equation}
T_{N} = A\exp(-B/x)
\label{eq.Tn}
\end {equation}
%%%%%%%%%%%%%%%%%%%%%%%%%%%%%%%%%%%%%%%%%%%%%%%%%%%%%%%
If we use this,  the best fitted value is $A = 2.3$ K and $B = 5.7
\times10^{-3}$. For the undoped sample (effectively 
$x = 2.3(2)\times10^{-4}$) $T_{N}$ 
is estimated to be 1 mK or lower, which is consistent with our failure to
observe an AF transition for the undoped sample.

\begin{figure}[t]
\begin{center}
\includegraphics*[width=7cm]{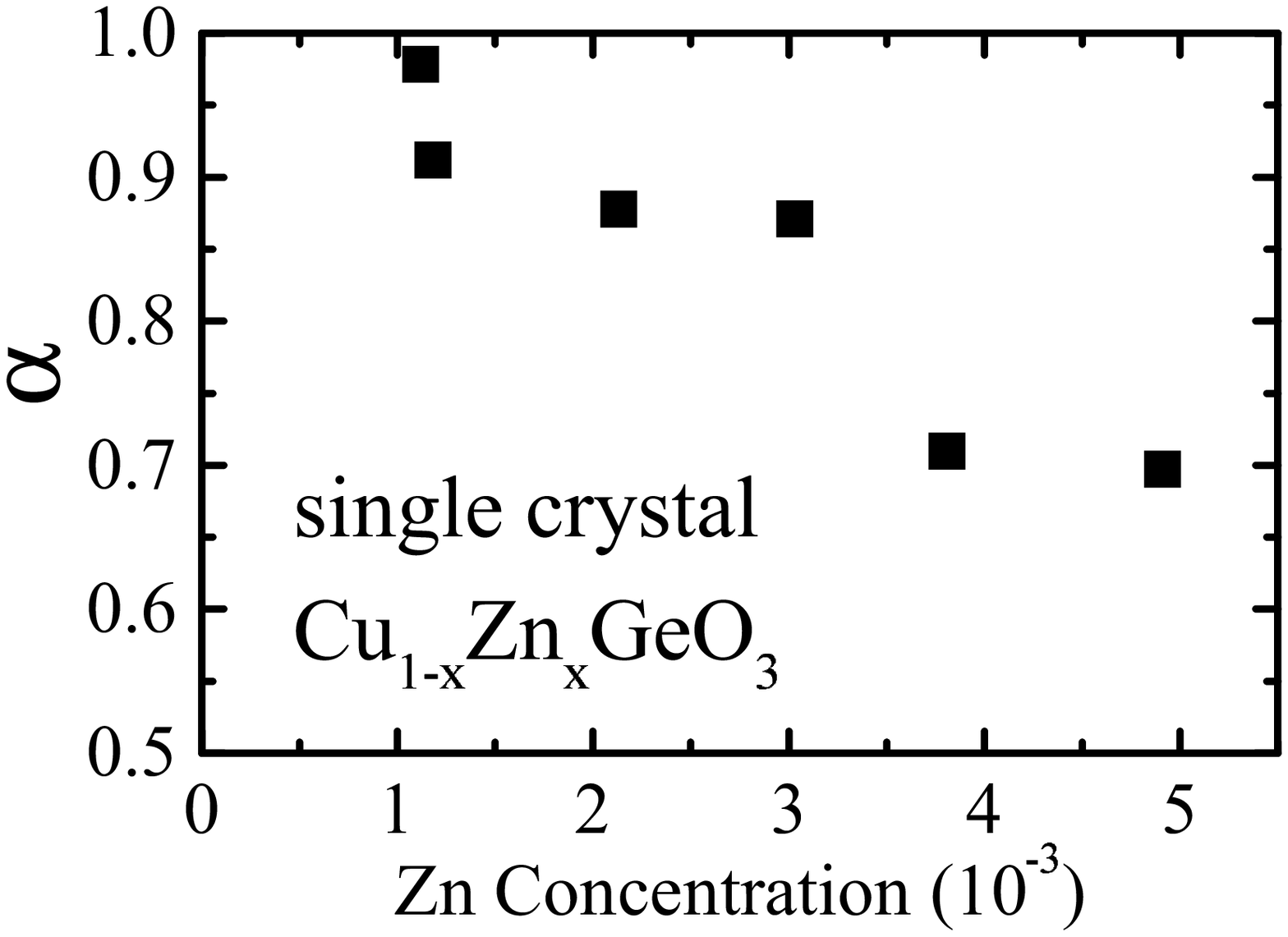}
\end{center}
\caption{%
The  constant $\alpha$ in Eq.~(\ref{eq.alpha}) as a function of 
Zn concentration $x$.
The definition of $\alpha$ is given in the text.
}%
\label{fig5}
\end{figure}

Now let us discuss the temperature dependence of the
susceptibilities above $T_N$.
As shown in Fig.~\ref{fig2}, all ${\chi}_c(T)$ at $T>T_N$
exhibit a weaker temperature dependence than $1/T$, which
indicates that there exists magnetic correlation between nearly isolated
$S=1/2$ states on the broken edges.
According to a theoretical calculation for a
1D-AF-Heisenberg-alternating chain with distributed AF
interaction \cite{Bulaevskii72} $J$, the susceptibility $\chi_c(T)$
at low temperatures is  given as
%%%%%%%%%%%%%%%%%%%%%%%%%%%%%%%%%%%%%%%%%%%%%%%%%%%%%%%
\begin{equation}
\chi = CT^{-\alpha},  ~~~~~       (0<\alpha < 1)
\label{eq.alpha}
\end {equation}
%%%%%%%%%%%%%%%%%%%%%%%%%%%%%%%%%%%%%%%%%%%%%%%%%%%%%%%
where $C$ and $\alpha$ are the constants which depend on the
distribution of $J$.
The value of $\alpha$ approaches to 1 as the magnetic correlation 
becomes weaker.
The best fitted value of $\alpha$ is given in Fig.~\ref{fig5} as a
function of Zn concentration $x$.
We can see a general trend that $\alpha$ increases with the
reduction of $x$.
For $x=1.12(2)\times10^{-3}$ where the
average distance between impurities is about 900$c$
($c$ is the lattice constant
along the chain, $c=2.94$ \AA),
$\alpha$ is still smaller than 1 ($\alpha = 0.978$).
If a coupling between these localized spins is
generated by the polarization of the spin-singlet background,
according to Ref.~\onlinecite{Fabrizio97}\ it is supposed to appear at
 %%%%%%%%%%%%%%%%%%%%%%%%%%%%%%%%%%%%%%%%%%%%%%%%%%%%%%%
\begin{equation}
T^{*} = T_{SP}\exp(-L/\xi)
\label{eq.T^{*}}
\end {equation}
%%%%%%%%%%%%%%%%%%%%%%%%%%%%%%%%%%%%%%%%%%%%%%%%%%%%%%%
where $L$ and $\xi$ are the average distance between Zn ions and
the soliton width, respectively.
When we use the calculated soliton width
$(\xi = 11.8 c)$ for the Si-doped CuGeO${}_3$,\cite{Fukuyama96a}\ %
$T^{*}$ for $x=1.12(2)\times10^{-3}$ is estimated to be
$1.1 \times 10^{-32}$ K.
This temperature is clearly much lower than the temperature range
 where the magnetic correlations between localized spins are observed.
 This fact implies that the magnetic correlations between
localized spins are enhanced by the relatively large interchain interaction
of CuGeO${}_3$ ($J_{b} = 0.1J_{c}$).

%%%%%%%%%%%%%%%%%%%%%%%%%%%%%
%%%%%%%%%%%%%%%%%%%%%%%%%%%%%%%%%%%%%%%%%%%%

Recently Masuda {\it et al.}\cite{Masuda98}\ reported that there
are two AF phases in Cu${}_{1-x}$Mg${}_x$GeO${}_3$.
One has the dimerization and AF-LRO at the same time at $x<x_c$ and
the other has only AF-LRO at $x>x_c$.
They called the former dimerized antiferromagnetic (D-AF) phase
and the latter uniform antiferromagnetic (U-AF) phase.
They found a first-order phase transition between the two phases as
the Mg concentration $x$ changes.
The critical concentration $x_c$ is about 0.023, where the SP
transition vanishes.
This fact shows that CuGeO${}_3$ would be a conventional (classical)
antiferromagnet, if there were no SP transition.
Apparently the dimerization suppresses the growth of the AF-LRO by
reducing the magnitude of the spin polarization.
On the other hand  the dimerization enhances the coherence of
the (AF) phase of the
spin polarization, because of the existence of the three-dimensional
LRO of the lattice (dimerization).
These competing effects of the dimerization seems to determine
the nature of the impurity-induced AF phase in low concentration region.
The introduction of impurities reduces the dimerization
and at the same time induces the spin polarization.
Growth of the polarization and the reduction of the dimerization
induce the AF-LRO at very low temperatures, which is the main reason why
we can observe the AF-LRO in such a low concentration region of
Cu${}_{1-x}$Zn${}_x$GeO${}_3$.

%%%%%%%%%%%%%%%%%%%%%%%%%%%%%
%%%%%%%%%%%%%%%%%%%%%%%%%%%%%%%%%%%%%%%

Until now it has not been so clearly determined
whether the same phenomenon occurs
in Zn-doped CuGeO${}_3$.
However it has been suggested that this phenomenon is universal
in the impurity-doped CuGeO${}_3$ at least when the substitution is done for
Cu${}^{2+}$, whether the impurities are nonmagnetic 
or magnetic (see the discussion in Ref.~\onlinecite{Masuda98}).
Apparently the samples studied in this paper belongs to D-AF phase
if there exist two AF phases in Zn-doped CuGeO${}_3$.

%%%%%%%%%%%%%%%%%%%%%%%%%%%%%%%%%%%%
Here we compare our work with the very recent one by Grenier
{\it et al.} on CuGe${}_{1-x}$Si${}_x$O${}_3$.\cite{Grenier98}
First we note that the systems are different: one is the substitution 
for Cu${}^{2+}$, which has $S=1/2$ spin on it, by Zn${}^{2+}$ ($S=0$)
and the other is the substitution for Ge${}^{4+}$ ($S=0$)
outside the spin chain by Si${}^{2+}$ ($S=0$).
However the reduction of the SP transition
and occurrence of AF-LRO have been observed to behave similarly and may
be compared.
As already stated, they measured the susceptibilities of
their samples with $x_{\mbox{\scriptsize Si}} \ge 0.002$ and $x=0$ 
down to 70 mK.
Their conclusion that the pure CuGeO${}_3$ does not undergo N\'eel
transition and there is no critical concentration for the occurrence of the
AF phase is consistent with our conclusion.
However if we take their scaling relation $x_{\mbox{\scriptsize Si}}
\simeq x_{\mbox{\scriptsize Zn}}/3$,
our minimum concentration sample
$x_{\mbox{\scriptsize Zn}}=1.12(2)\times 10^{-3}$
corresponds to $x_{\mbox{\scriptsize Si}}\simeq 4\times 10^{-4}$,
which is about 5 times less
than that of Ref.~\onlinecite{Grenier98} and in fact this sample shows
N\'eel transition at $T_N=0.0285$ K, which is one order of magnitude
lower than that of
theirs ($T_N=0.25$ K).
Moreover their values of the concentration in the very low concentration
region are uncertain and they drew their conclusion (absence of
critical concentration) by modifying the
concentration values by fitting to the suppression of SP transition (see
Ref.~\onlinecite{Grenier98}).
This suggests (and also they themselves
stated) that Zn-doped CuGeO${}_3$ is more suitable for this kind of study.
On the pure CuGeO${}_3$ sample we measured the susceptibility down to
5 mK compared to their 70 mK.
Therefore we may say that our study is extensive on the study of
AF-LRO in the very low concentration region and that
our experimental results and conclusion are more firmly
established than theirs.
Another important point is, as already stated, Zn-doped and Si-doped
CuGeO${}_3$ may belong to different systems.
The phase diagram obtained in
Fig.~8 of Ref.~\onlinecite{Grenier98}\ seems to indicate that
there is no first-order
phase transition between D-AF and U-AF phases in Si-doped CuGeO${}_3$ in
contrast to the observation of the one in Mg-doped CuGeO${}_3$ by Masuda
{\it et al.}\cite{Masuda98}
This casts a grave doubt on the belief that the AF phase(s) in the both
substitution systems for Cu${}^{2+}$ and for Ge${}^{4+}$ is similar.
Therefore it is highly desirable to do the detailed experiments in
both systems and compare them.

In conclusion we have revealed how the AF-LRO behaves when
the Zn concentration is extremely low.
The undoped CuGeO${}_3$ does not have AF-LRO down to 12 mK.
On the other hand, all the Zn-doped samples showed AF-LRO.
The lowest observed $T_N$ was 0.0285 K in the sample with
$x=1.12(2)\times 10^{-3}$.
From the concentration dependence of $T_N$ we concluded the absence of a
critical concentration for the occurrence of AF-LRO.
This may come from the fact that the dimerization sustains
the phase coherence of the spin polarization and
is also consistent with the theory
of the impurity-doped SP system.\cite{Fukuyama96a}
The temperature dependence of the susceptibilities
at $T>T_N$  indicates that the magnetic correlations between
localized spins are enhanced by
a relatively large interchain interaction
of CuGeO${}_3$.

This work was supported in part by Grant-in-Aid for Scientific
Research (A),
Grant-in-Aid for Scientific Research on Priority Area ``Mott transition'',
Grant-in-Aid for COE Research
``Spin-Charge-Photon'',
from the Ministry of Education, Science, Sports, and Culture,
and NEDO International Joint Research Grant.

% FIGURES CAPTIONS %

%%%%%%%%%%%%%%%%%%%%%%%%%%%%%%

\end{multicols}

\end{document}